# Atomic-scale imaging of the surface dipole distribution of stepped surfaces

Carmen Pérez León,*,† Holger Drees,† Stefan Martin Wippermann,‡ Michael Marz,*,† and Regina Hoffmann-Vogel†

†Physikalisches Institut, Karlsruhe Institute of Technology (KIT), Wolfgang-Gaede-Str. 1, D-76131 Karlsruhe, Germany, and ‡Max-Planck-Institut für Eisenforschung GmbH, Max-Planck-Straße 1, D-40237 Düsseldorf, Germany

E-mail: cperezleon.science@gmail.com; michael.marz@kit.edu

**ABSTRACT:** Stepped well-ordered semiconductor surfaces are important as nanotemplates for the fabrication of one-dimensional nanostructures. Therefore a detailed understanding of the underlying stepped substrates is crucial for advances in this field. Although measurements of step edges are challenging for scanning force microscopy (SFM), here we present simultaneous atomically resolved SFM and Kelvin probe force microscopy (KPFM) images of a silicon vicinal surface. We find that the local contact potential difference is large at the bottom of the steps and at the restatoms on the terraces, whereas it drops at the upper part of the steps and at the adatoms on the terraces. For the interpretation of the data we performed density functional theory (DFT) calculations of the surface dipole distribution. The DFT images accurately reproduce the experiments even without including the tip in the calculations. This underlines that the high-resolution KPFM images are closely related to intrinsic properties of the surface and not only to tip-surface interactions.

Stepped well-ordered surfaces and, in particular vicinal semiconductor surfaces, are well suited for applications as nanotemplates for the fabrication of one-dimensional nanostructures.[1-7] Among such structures, monoatomic wires are candidates of interesting electronic properties such as Luttinger-liquid behavior.[8] The vicinal Si(111) surface with 10° miscut towards the [$\bar{1}\bar{1}2$] direction is a popular stepped surface that can be used as a model system. This surface contains (7×7) reconstructed terraces oriented along the Si(111) direction, a well characterized and understood surface, separated by a stepped region. The presence of the (7 × 7) reconstructed areas makes this vicinal system an ideal testbed for surface characterization techniques and investigating its rich morphology and electronic features. Teys et al. proposed that this surface is oriented along the (7 7 10) direction.[2] Within Teys model, the stepped part consists of a periodically ordered triple step with a height of 3 atomic layers and a width of 16 atomic rows, corresponding to a lateral periodicity of 5.2 nm.

Scanning force microscopy (SFM), particularly in ultra-high vacuum (UHV) and its non-contact mode, has become one of the standard techniques for analyzing the topographic properties of flat surfaces at the atomic scale.[9] On conducting surfaces, SFM provides complementary information to that obtained with scanning tunneling microscopy (STM),[10,11] in some cases with even higher spatial resolution.[12,13] The atomic resolution capability of SFM arises from the short-range forces acting between an atomically sharp tip and a clean surface.[14] The differences between the work functions of the probing tip and the surface of the sample give rise to contact potential differences (CPD) that can be measured using Kelvin probe force microscopy (KPFM).[15-19] The origin of atomic-scale KPFM contrast is still under discussion, since the work function is considered as a macroscopic concept and each material surface should have a given value. A recent work on Si(111)-(7 × 7), where experiments were compared to calculations, found that the modulation of the surface dipole affected by the presence of the tip could explain the atomic-scale contrast.[17] So far these studies have been restricted to flat surfaces, because scanning over a corrugated surface can lead to strong interaction between atoms of the tip apex with the surface, and thus to a loss of resolution.

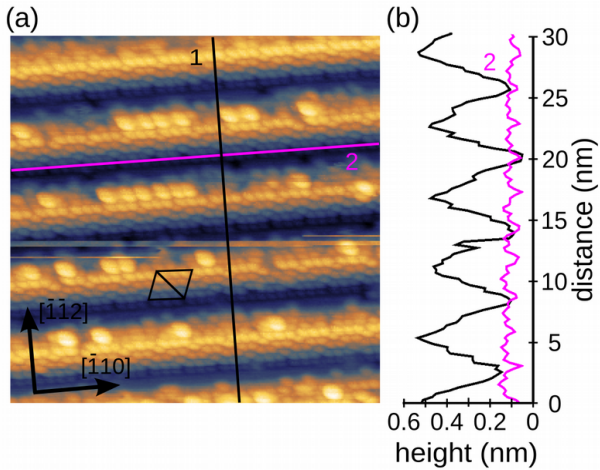

Figure 1: FM-SFM images of the clean vicinal Si(111) surface at RT with atomic resolution. (a) Large terrace consisting of periodically spaced steps and Si(111)-(7 × 7) reconstructed areas. Image size 30×30 nm$^2$. A Si(111)-(7×7) unit cell is indicated by triangles. (b) Line profiles of the linecuts on (a) displaying the height and periodicity of the steps and (7 × 7) reconstructed terraces.

Here, we present atomically-resolved frequency modulated SFM (FM-SFM) and KPFM images of the vicinal Si(111) surface. Our high-resolution images supported by density functional theory (DFT) calculations elucidate the origin of the Kelvin contrast at the triple step and the influence of the tip-surface distance.

Figure 1(a) shows an atomically-resolved FM-SFM image of the silicon surface after preparation. The surface consists of large terraces which contain periodically spaced steps and Si(111)-(7 × 7) reconstructed areas. In general, we avoid scanning with the fast axis parallel to the step edges by rotating the scan direction by 5°. Still, sometimes the tip apex gets unstable, e.g. in the middle of the image where a jump occurred in the stepped part. Large protrusions are observed on the step edges of the (7 × 7) reconstructed terraces. We tentatively ascribe them to additional silicon clusters produced during the preparation of the sample. In Fig. 1(b), profiles perpendicular and parallel to the step edges display the surface corrugation and periodicity of the (7 × 7) reconstructed areas, respectively.

In order to discuss the structure of the surface, 3D and 2D SFM images are plotted in Fig. 2. We use the notation introduced by Teys and co-workers.[2] In the images, a (7 × 7)-reconstructed Si(111) terrace is followed by a triple step and the next (7 × 7) terrace. The triple step ($S_1$, $S_2$, $S_3$) consists of several adatoms and dimer rows. $S_3$: On the edge of the upper terrace, the last row of silicon adatoms (denoted as $A_3$, the index indicating the layer) is accompanied by a row of parallel dimers ($D^\parallel_3$). $S_2$: The lower layer (index 2) is formed by a row of adatoms ($A_2$). Often atomic defects are observed in the $A_2$ row at the places where corner vacancies of the (7 × 7) surface and no $D^\parallel_3$ dimers are found (these defects are denoted as RD). Such RD appear due to the mismatch between the 2-fold periodicity of the dimer rows and the 7-fold periodicity of the terraces. At the RD, the adatom is shifted towards the step, as indicated in Fig. 2(b). In Fig. 2(c), profiles over a defective and non-defective step line of Fig. 2(b) are shown, and the place of a RD is indicated. Below the row of $A_2$, we found rows of perpendicular and parallel dimers ($D^\perp$, $D^\parallel_1$, respectively). Many dimers are missing in the FM-SFM image, especially in $D^\parallel_1$. $S_1$: Below the dimers, two extra rows of silicon atoms are observed, the ZR row, reported from STM analysis to have a zigzag structure,[2] and the R row located at the bottom of the triple step. The presented SFM image coincides almost one to one with the empty-states STM image obtained by Teys et al.[2] In the case of silicon, the empty states are the fingerprint of the free dangling bonds. This supports that the SFM image arises from covalent bonding of the tip apex with the surface atoms. For a more detailed SFM and DFT analysis of the atomic structure of the triple step, and a rigorous comparison with the STM data cf. Ref. 20.

Fig. 3 displays simultaneous topographic FM-SFM and KPFM images measured over a terrace, the triple step, and the next terrace. In the topographic image, Fig. 3(a), several defects are observed at step edges that we previously attributed to Si clusters. Furthermore, many atoms are missing in the $D^\parallel_3$ and $A_2$ rows. In the Kelvin image, Fig. 3(b), strong surface potential variations on the terraces and the step are observed. Considering the background local CPD (LCPD) obtained at positions between adatoms and at the bottom of the step as the reference value (ca. 0.35 V), we discuss the values of the LCPD as higher or lower than this reference. Thus, the adatoms of the (7 × 7) terraces have the lowest LCPD, 0.1 − 0.2 V in our image, in accordance with previous reports on flat Si(111).[17,19] The defects at step edges (Si clusters) have a similar LCPD to the adatoms, supporting that they are composed of additional Si atoms. The R row and $A_2$ adatoms also have a low LCPD compared to the reference, but slightly higher than the adatoms and clusters. The ZR row shows almost no contrast difference with respect to the background. On the corner holes, on the other hand, the LCPD appears slightly higher than the background. Also in the region between the adatoms of the (7 × 7) terraces (restatoms), a higher LCPD is imaged (≥ 0.4 V). Remarkably, the region of the dimers, in particular at $D^\parallel_1$ and $D^\perp$, show a high LCPD (ca. 0.5 V).

Over the flat (7 × 7)-reconstructed terraces, the general contrast and the absolute value of the local contact potential difference is in agreement with previous KPFM studies also performed together with FM-SFM.[17] Sadewasser et al. considered several factors as possible origin of the atomic contrast: a possible tunneling current flow between tip and surface, the modulation of the surface dipole, and the influence of the formation of chemical bonds between the tip apex and surface atoms.[17]

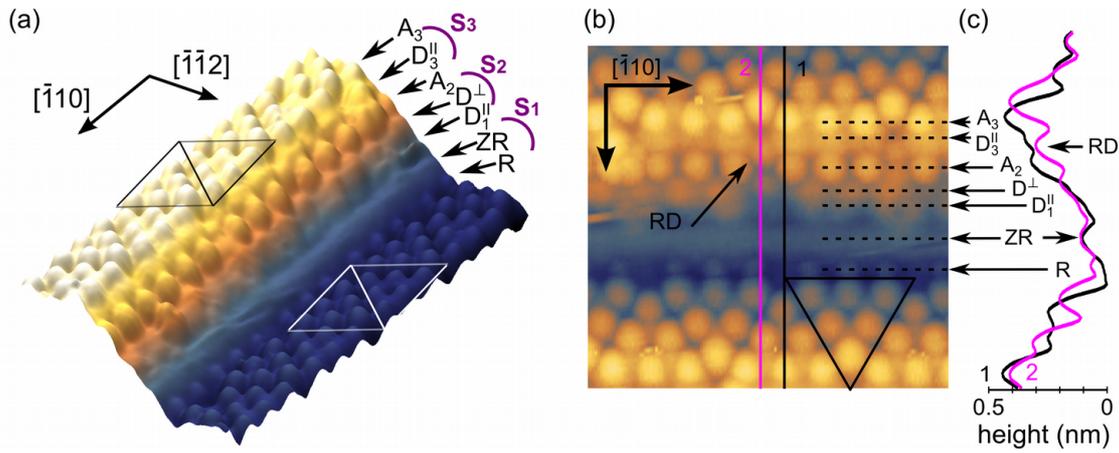

Figure 2: FM-SFM images of the structure of the vicinal Si(111) surface at RT. (a) 3D and (b) 2D plots. (c) Line profiles of the linecuts on (b) unveil the crystal lattice cross-section. The surface consists of flat (111)-(7 × 7) reconstructed terraces, followed by a triple step ($S_1$, $S_2$, $S_3$) formed by adatoms ($A_2$, ZR, R) and dimers ($D^\parallel_3$, $D^\perp$, $D^\parallel_1$). Half cells are marked with triangles.

We analyze first the influence produced by a tunneling current flow between tip and surface. The tunneling probability is different at different atomic sites (as reported from STM images[2]). This may affect the strength of the LCPD, but we believe this effect to be small since the current flow is below the detection limit of our experimental setup.

We continue with the work function that is generally determined by the surface dipole, which reflects the charge distribution on the surface.[17,21] In order to better understand the Kelvin contrast, we performed DFT calculations of the electrostatic surface potential. The tip is excluded for obtaining directly information on the physical properties of the surface and not on the deviations caused by the interaction of the surface with the tip. We define the local work function $\Phi_{loc}$ at a point r near the surface as $\Phi_{loc}(\mathbf{r}) = V_{eff}(\mathbf{r}) - E_F$, where $V_{eff}(\mathbf{r})$ is the single-particle effective electrostatic potential and $E_F$ is the Fermi energy.[22] The calculated $\Phi_{loc}$ is shown in Fig. 3(c). The overall behaviour of the local work function along the surface is remarkably well reproduced by the simulations. For the discussion of the data, we divide the surface into two parts: the well-known (7 × 7) reconstructed terraces and the triple step.

The local atomic dipole distribution on the flat Si(111)-(7 × 7) surface was investigated by Cho and Hirose.[23] They detected dipole moments pointing upwards on the Si adatoms, whereas the dipole moments pointed inwards at the interstitial sites between Si adatoms.[23] This behavior was also observed in Ref. [17] and is consistent with the results of this work. We can relate these dipoles to the different charge distribution on the different atoms, i.e., at the lower-lying restatoms the partial negative charge is larger than at the more protruding adatoms.

At the triple step we find a striking agreement between our experimental results and our simulations. The upper part of every step of the triple step displays a lower electron density (i.e. lower LCPD), while the bottom of the steps shows a higher one. The DFT calculations (Fig. 3(c)) reveal that the high electron density is not only associated to the dimers but also to the Si restatoms that lie in the lowest part of every step. At the upper part of the triple step, close to the $D^\parallel_3$ dimer row, the electron density is lower. At the bottom of this step ($S_3$), located between two RDs and between the $D^\parallel_3$ dimers and $A_2$ adatoms, there are two restatoms denoted as $RA_2$. These appear as bright features in the experiment and simulations. The $A_2$ adatoms at the upper part of the second step ($S_2$) show a lower $\Phi_{loc}$, while between $D^\perp$ and $D^\parallel_1$ at the bottom display a high one. Again at the upper part of the lower step ($S_1$), between $D^\parallel_1$ and ZR, the electron density is lower. Finally, at the bottom of $S_1$ between ZR and R, the electron density is high. There, we find additional restatoms, denoted as RB, which are the deepest located atoms within the reconstruction with dangling bonds. Further details of the discussion of the DFT data have been included in the Supporting Information.

One important difference between our simulations and experimental results is that at the bottom of the triple step, at RB, the partial negative charge appears much larger in the calculations than in the Kelvin image. Due to the presence of steps, the overall distance between tip and sample varies strongly in the experiment during the scan, compared to the measurements over a flat surface, see profile in Fig. 2(c). The tip-surface distance at which the potential is calculated (0.325 nm), is much closer than the true height of the tip in the experiment. Adding a correction to this height profile to the simulation of the surface dipole on this sample area, results in a more diffuse and less bright feature in closer agreement with the Kelvin image, as shown in Fig. 4.

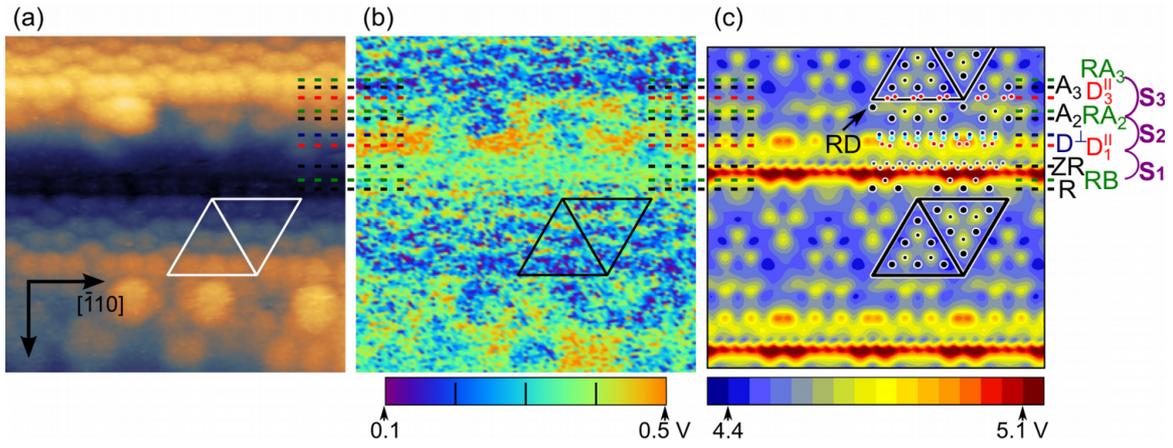

Figure 3: Atomically resolved (a) FM-SFM and (b) simultaneously obtained Kelvin image at RT. (c) Calculated local work function from the electrostatic effective single-particle potential. The local contact potential difference is large at the bottom of the steps and at the restatoms on the terraces, whereas it drops at the upper part of the steps and at the adatoms on the terraces.

The observed features are comparable to the Smoluchowski effect known for metallic surfaces.[24] Smoluchowski stated that the electronic density at step edges follows the step edge more smoothly than expected from prolonging the bulk electronic density, creating a dipole pointing upwards. Here, we directly image the different charges of such dipoles, which gives rise to an alternate variation of charge and therefore of Kelvin contrast. A scheme of the charge distribution is presented in Fig. 5.

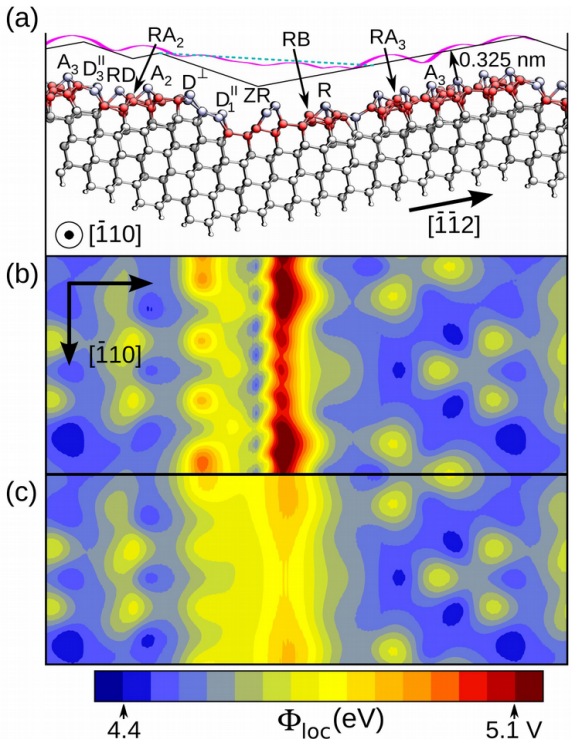

Figure 4: (a) Schematic side view of the Si(7 7 10)-(16×14) model of the structure. The purple curve represents the experimental line profile (Fig. 2(c)), adjusted to scale. The black solid and turquoise dashed lines indicate the height profiles at which the local work functions $\Phi_{loc}$ in (b) and (c) have been calculated, respectively. The turquoise dashed line is closer to the measured line profile.

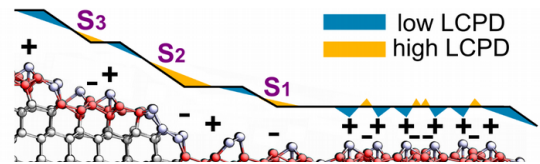

Figure 5: Scheme of the charge distribution at the triple step and the following (7 × 7) reconstructed terrace.

Finally, we analyze the influence of the formation of chemical bonds between the tip apex and surface atoms. In Ref. [17] bond formation was reported to induce a local redistribution of the charge density leading to a change of the surface dipole and, consequently, to variations of the local chemical potential. A strong redistribution was identified for close tip-surface separations, whereas this was reduced for larger ones.[17] In Fig. 4(a), we observe that at the stepped region the height of the tip in our experiment is higher than at the flat part, pointing to a weaker effect of this bonding formation in the KPFM images in this region. This argument reinforces the close agreement of experiment and simulations even without taking into account the effect of the tip in the calculations. This agreement is not only qualitative but also quantitative. The differences between the work function at the adatoms and restatoms is around 0.5 − 0.6 V in the calculations, very close to the 0.4 V in the experiments.

Summarizing, we demonstrate that FM-SFM and KPFM are able to provide simultaneous atomic resolution of the topography and the surface potential distribution even at stepped surfaces. For a better understanding of the KPFM image, we performed DFT calculations of the electrostatic surface potential obtaining a striking agreement with the experiments, even without including the influence of the tip. We observe a large LCPD at the restatoms whereas it drops at the adatoms. We explain the origin of these features in terms of the local environment of the atoms and partial charge transfers. This explanation is general and not unique to Si surfaces, which serves as model system. The DFT calculations uncover the nonnegligible contribution of the different restatoms on the surface to the surface dipole, and the influence of the tip-surface distance. These results demonstrate that the observed KPFM atomic contrast is not only produced by the interaction with the tip, but indeed reflects an intrinsic property of the surface.

## SUPPORTING INFORMATION
Experimental methods. Numerical methods. Details of the DFT calculations of the surface dipole.

## ACKNOWLEDGMENT
We thank the European Research Council for financial support through the starting grant NANOCONTACTS (No. ERC 2009-Stg 239838), and O. Stetsovych and O. Custance for discussions about SFM methods. S. W. acknowledges BMBF NanoMatFutur grant No. 13N12972 from the German Federal Ministry for Education and Research, and thanks W. G. Schmidt for helpful discussions.

# Supporting Information:
# Atomic-scale imaging of the surface dipole distribution of stepped surfaces

Carmen Pérez León,*,† Holger Drees,† Stefan Martin Wippermann,‡ Michael Marz,*,† and Regina Hoffmann-Vogel†

†Physikalisches Institut, Karlsruhe Institute of Technology (KIT), Wolfgang-Gaede-Str. 1, D-76131 Karlsruhe, Germany, and ‡Max-Planck-Institut für Eisenforschung GmbH, Max-Planck-Straße 1, D-40237 Düsseldorf, Germany

E-mail: carmen.perez.leon@kit.edu; michael.marz@kit.edu

**Contents:**

**A- Experimental methods**

**B- Numerical methods**

**C- Details of the DFT calculations of the surface dipole**

## A- Experimental methods

Both sample preparation and experiments were carried out in an ultra-high vacuum (UHV) chamber with a base pressure of less than $3 \cdot 10^{-8}$ Pa. Stripes of Si(111) low n-doped (phosphorus, $\rho = 1 - 10$ Ωcm, Virginia Semiconductor) and an inclination angle of $10 \pm 0.5°$ towards the $[\bar{1}\bar{1}2]$ direction were used. The silicon stripes were cleaned in an diluted aqueous HF solution prior to loading into the UHV chamber. The sample was resistively heated by direct current with the current direction parallel to the steps on the vicinal Si(111). The surface was prepared by several short flashes to 1420 K, the last flash was proceeded by a fast ramp down to 1200 K, followed by a slower cooldown. The silicon sample was then transferred to a variable-temperature scanning force microscope (Omicron NanoTechnology GmbH, Taunusstein, Germany) equipped with Nanosensors cantilevers (Neuchatel, Switzerland) and a Nanonis phase-locked loop electronics (SPECS, Zurich, Switzerland). All measurements were performed in the non-contact mode. Topographical imaging was carried out at constant frequency shift using sputtered silicon cantilevers with a force constant of $30 - 50$ N/m, and a free resonance frequency of $270 - 300$ kHz. Some of the cantilevers used were coated with Platin-Iridium. The topography images were obtained while applying a voltage that compensated the local contact potential difference (LCPD) between the tip and the sample. When KPFM measurements were performed in parallel to the topography measurements, an ac-voltage was applied to the tip at an oscillation frequency of 619 Hz and an amplitude of 0.7 V. For characterizing the frequency modulated scanning force microscopy images the normalized frequency shift ($\gamma = \Delta f \cdot k \cdot A^{3/2} /f_0$) has been used. For the discussion of the structure, the experimental drift has been compensated in some of the presented images. All measurements were performed at room temperature.

**Imaging parameters of figures in main article**

Fig. 1: $\Delta f = -16$ Hz, A = 7 nm, k = 32 N/m, $f_0$ = 295 KHz, $\gamma = -1$ fN$\sqrt{m}$. Si cantilever tip.

Fig. 2: $\Delta f = -60$ Hz, A = 8 nm, k = 46 N/m, $f_0$ = 272 KHz, $\gamma = -7.3$ fN$\sqrt{m}$. Pt-Ir coated cantilever tip.

Fig. 3: $\Delta f = -17$ Hz, A = 8 nm, k = 32 N/m, $f_0$ = 295 KHz, $\gamma = -1.3$ fN$\sqrt{m}$. Si cantilever tip.

## B- Numerical methods

The electronic structure calculations were performed within density functional theory (DFT)and the Purdue, Burke, Ernzerhof (PBE) generalized gradient approximation,[1] as implemented in the Vienna ab initio Simulation Package (VASP).[2] For the Brillouin Zone (BZ) integrations in the electronic structure calculations uniform meshes equivalent to 224 points for the (1×1) surface unit cell were used. The starting structures were prepared based on the experimental observation of a 16-fold lateral periodicity, the model introduced by Teys et al.,[3] and variations thereof. Due to the mismatch between the 7-fold periodicity of the terraces and the 2-fold periodicity of the dimer rows, the smallest possible unit cell size parallel to the step edges that does not contain any obvious defects, e.g. $D^{\parallel}_1$ monomers instead of dimers, has a 14-fold periodicity. Thus, all calculations were performed within a (16×14) surface cell. The surface has been modeled by a slab containing four Si bilayers along the (111) direction, resulting in ~2300 atoms per unit cell. The bottom Si bilayer was frozen at the equilibrium DFT lattice constant with the dangling bonds terminated by hydrogen. To avoid a spurious interaction between periodic images along the surface normal, a vacuum distance of 50 Å between the surface and the bottom layer of its periodic image has been employed. This vacuum distance has been used in conjunction with a dipole correction by introducing a step discontinuity inside the vacuum region which cancels out the surface dipole. Forces were relaxed below a threshold of 0.01 eV/Å.

## C- Details of the DFT calculations of the surface dipole

Due to the mismatch between the 2-fold periodicity of the dimer rows and the 7-fold periodicity of the terraces, row defects (RDs) appear within the $A_2$ adatom row. These RDs passivate the otherwise dangling bonds of the Si atoms at the corner hole. There are no such RD at the corner holes at the opposite side of the (7×7) terrace, this asymmetry produces a difference in the LCPD between both types of corner holes.

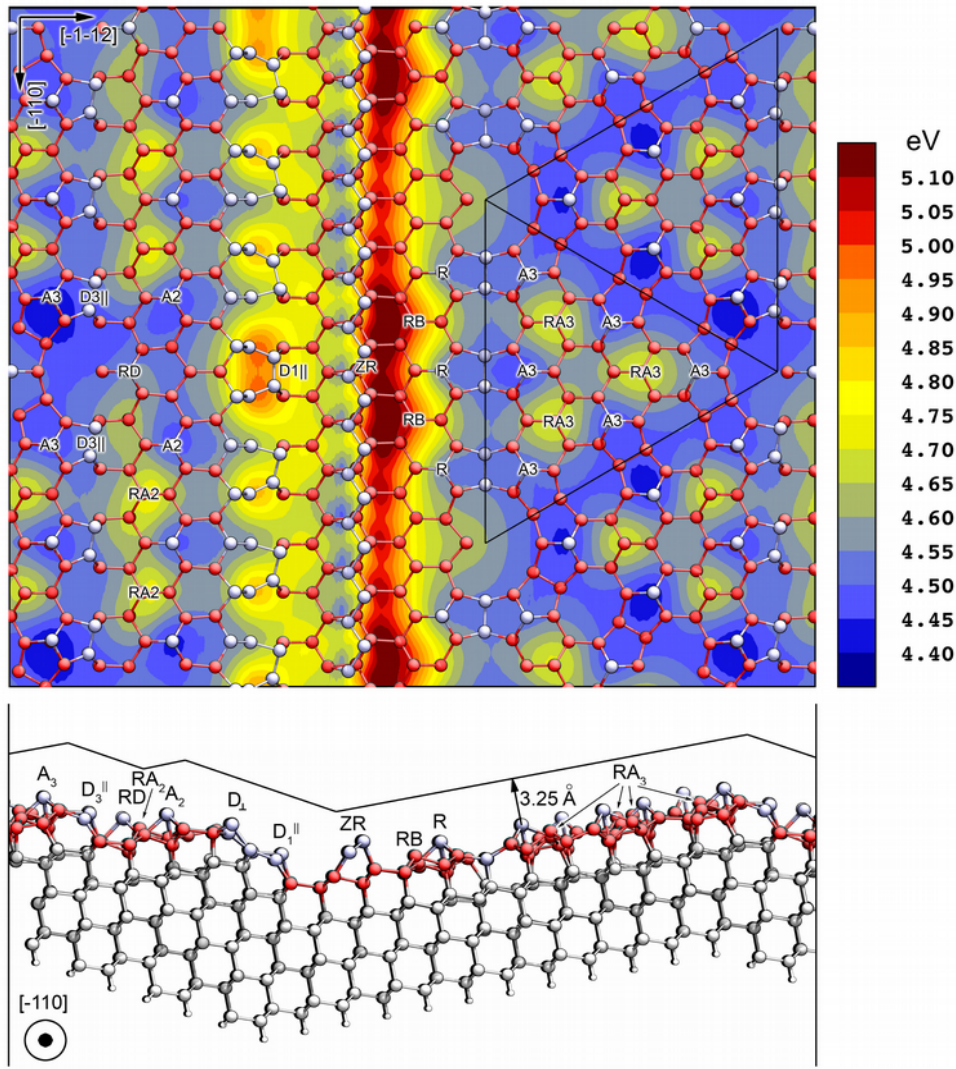

Located between two RDs and between the $D^\parallel_3$ dimers and $A_2$ adatoms, there are two rest atoms denoted $RA_2$. They appear as bright features in the simulations. This is due to the shift of electronic density towards these rest atoms because of their location at the bottom of the first part of the triple step. Both, the step edge and the $A_2$ adatoms, contribute to the charge transfer towards $RA_2$.

The $D^\perp$ dimer row is shown in our DFT calculations to prefer a buckled configuration, where the $D^\parallel_1$ dimers assume an alternating more upright and more flat orientation. $D^\parallel_1$ dimers in the upright orientation feature a slightly higher local work function $\Phi_{loc}$, introducing a 2-fold modulation along the $D^{\parallel}_1$ row. Within this dimer row, there are defects intrinsic to the reconstruction where two adjacent $D^{\parallel}_1$ are both oriented in the upright configuration. These defects occur regularly to accommodate the mismatch between the 7-fold symmetry of the 7×7 reconstructed part and the dimer row with a 2-fold symmetry. They appear particularly bright in the simulation and explain why in experiment some of the $D^{\parallel}_1$ dimers exhibit a high and some a low LCPD, respectively. Presently, the precise reason why the small differences in the orientation in the dimer row cause such noticeable differences in $\Phi_{loc}$ remains an open question. Additionally, there are extra Si-adatoms adsorbed at the step edges. This makes it difficult to distinguish experimentally between LCPD differences caused by either dimer orientation or extra Si-adatoms, due to the inherently reduced resolution directly at the step.

The ZR row is preferentially tilted away from the step edge in our calculations. However, an equal tilt in the other direction represents another local energy minimum. Even tilts in different directions of different ZR row parts are relatively stable. Movements of the ZR row induced by the tip may contribute to the smeared appearance of ZR in the experimental results.